\documentstyle[aps,prl,twocolumn]{revtex}
\begin{document}
\title{N\'eel state of antiferromagnet as a result of a local
  measurement in the distributed quantum system}
\author{M. I. Katsnelson \cite{mik}, V. V. Dobrovitski, and 
  B. N. Harmon}
\address{Ames Laboratory, Iowa State University, Department of
  Physics and Astronomy, Ames, Iowa, 50011}
\date{today}

\maketitle
\draft
\begin{abstract}
Single-site measurement in a distributed macroscopic
antiferromagnet is considered; we show that it can create
antiferromagnetic sublattices at macroscopic scale. We
demonstrate that the result of measurement depends on the
symmetry of the ground state: for the easy-axis case the N\'eel
state is formed, while for the easy-plane case unusual ``fan''
sublattices appear with unbroken rotational symmetry, and a 
decoherence wave is generated. For the latter case, a 
macroscopically large number of measurements is needed to 
pin down the orientation of the sublattices, in spite of the
high degeneracy of the ground state. We 
note that the type of the final state and the appearance of
the decoherence wave are governed by the degree of entanglement 
of spins in the system.
\end{abstract}


The ordered antiferromagnetic (AFM) state has been extensively
discussed since the experimental discovery of
antiferromagnetism \cite{neel,astrov}. The
conventional picture, associated with L.\ N\'eel, represents a
macroscopic antiferromagnet as consisting of two sublattices,
each possessing a macroscopically large total spin ${\bf S}_A$.
This picture agrees with the experiments on macroscopic
antiferromagnets \cite{astrov}, but serious conceptual problems
arise. The ground state of a Heisenberg magnet with AFM exchange
interactions should be a nondegenerate singlet with zero total
magnetization of each sublattice. Moreover, the AFM ground state
should be time-reversal invariant, while the N\'eel state is
obviously not. 
For a long time, it was commonly believed that the
difficulty can be resolved by recognizing that due to very high
(quasi)degeneracy of the AFM ground state (the separation 
between
the lowest levels is of order $1/N$, where $N$ is the total
number of spins in the sample), ``extremely weak forces can pin
down the orientation'' (quote from Ref.\ \cite{pwa}) of the
sublattices thus breaking the rotational and time-reversal
symmetries and leading to the N\'eel state. Latter advances in 
quantum measurement theory \cite{zurek} demonstrated that 
it is not the interaction strength itself that
matters, but its ability to build up quantum correlations
(entanglement) between the system and its environment,
so that the system decoheres and the initial symmetry breaks.
In many realistic situations, the process of decoherence is
extremely fast, and can be considered as a quantum measurement 
in the von Neumann's sense \cite{vonneu}, i.e. as an 
instantaneous projection of the system's initial state onto 
the final state with broken symmetry.

However, in the present paper, we show that building the argument
upon quasidegeneracy is intrinsically misleading. We present two
situations, realistically modeling easy-plane and easy-axis
antiferromagnets, which have the same quasidegeneracy (with the
lowest levels separation $1/N$), and we show that the impact of 
environmental
measurement on each of the systems is drastically different:
although time-reversal symmetry breaks in both cases, this is 
not so for the rotational symmetry, i.e.
sublattices can emerge without formation of the N\'eel state,
leaving the orientation of sublattices magnetization 
indeterminate. We show that the different scenarios take place
because of different degrees of entanglement between the
individual spins in an antiferromagnet.

Appearance of the AFM sublattices due to environmental
``measurement'' has been studied \cite{stamp} for the case
of very small antiferromagnetic particles using the method of
collective coordinate, by mapping of the 
low-energy part of the antiferromagnet's spectrum onto a 
two-state system \cite{twostate}. Such a mapping is valid when
all the excited states are well-separated from the doublet of
lowest levels, and the non-collective dynamics can be neglected,
i.e. in the case of easy-axis antiferromagnet with noticeable 
anisotropy.
But in macroscopic samples the well-separated ground
doublet can be absent (e.g., the long-wavelength magnons in the
easy-plane antiferromagnet are gapless, and corresponding
excitations can be very close to the ground state), and the
mapping onto a two-state system is not legitimate. Also,
measurement in a distributed quantum system often generates a
decoherence wave \cite{decwav} which can not be described by the
collective coordinate method. Furthermore, in reality, the
interactions with dissipative degrees of freedom, such as
nuclear spins, are short-ranged, so that corresponding
measurements are essentially {\it local}. Locality 
of the measurement, which discriminates the given site
(belonging, to the sublattice A) from its neighbors (belonging to
the sublattice B), is crucial; the global measurement which does
not distinguish sites from different sublattices can not
resolve the N\'eel state and can not cause the formation of
sublattices \cite{irkat}.

Therefore, to describe the appearance of sublattices, we have to
consider {\it local} measurement in a {\it distributed} quantum
spin system. To our knowledge, this problem has never been
applied before to the study of antiferromagnets. In this paper,
we show that even the microscopic single-site measurement can
create sublattices in the macroscopically large antiferromagnet
which was initially in the time-reversal-symmetric state. 
We study this process in
detail and show that it can be very different from what is 
suggested by the collective coordinate method or similar models.

Let us consider the Heisenberg antiferromagnet containing $N$ 
sites with the spin $S$ at each site; the isotropic exchange part
of its Hamiltonian is
\begin{equation}
\label{ham0}
{\cal H}_0 = \sum_{\bf q} J_{\bf q} (S^+_{\bf q}S^-_{\bf q} +
  S^z_{\bf q}S^z_{\bf q}),
\end{equation}
where $S^+_{\bf q} = (1/\sqrt{N}) \sum_j s^+_j
\exp{(i{\bf q r}_j)}$, and, similarly, $S^-_{\bf q}$
and $S^z_{\bf q}$, are the Fourier transforms of the
single-site spin operators: $s^+_j$
and $s^-_j$ are the ladder spin operators for the site
$j$, and $s^z_j$ is the operator of the $z$-projection
of the spin at the site $j$. Analogously, $J_{\bf q}$
is the Fourier transform of the exchange integral,
where $\sum_{\bf q} J_{\bf q} = 0$.
If the exchange integrals are such that 
$\max_{\bf q} J_{\bf q} = J_{\bf Q}$ and $2{\bf Q}$ coincides
with the vector of the reciprocal lattice, then the AFM ordering
appears along the vector $\bf Q$; we assume infinite correlation
length since the finite-temperature effects are beyond the scope 
of this paper.

As will be shown below, it is important to take into account
the symmetry of the ground state imposed by the anisotropic
interactions, so we assume a very small anisotropy
term ${\cal H}'$, e.g., of the form
${\cal H}' = D \sum_j (s^z_j)^2$,
also present in the system's Hamiltonian.
It can be seen that for very small $D$ the analysis given below
does not depend on the specific form of the anisotropy, 
nor on its magnitude: symmetry of the ground state is 
predominant, so we assume $D\to \pm 0$ below. 

First, let us consider the case of easy-plane anisotropy,
$D\to +0$. Exact eigenstates of the system are not
known, except for some very special situations.
To describe AFM states in the general case, the following 
family of time-reversal-symmetric (TRS) states has been 
proposed \cite{irkat}:
\begin{equation}
\label{trsep}
|\Psi_M\rangle =\left[\frac{N^M (2NS-M)!}{M! (2NS)!}\right]^{1/2}
  \left(S^-_{\bf Q}\right)^M |0\rangle
\end{equation}
where $M=0,\dots,2NS$ is an integer, and $|0\rangle$ is the 
ferromagnetic state (all spins up).
The state $|\Psi_M\rangle$ is an eigenstate of the 
$z$-projection of the operator of the total system's spin with 
the eigenvalue ${\cal S}_z = NS-M$, so the AFM ground state 
corresponds to $M=NS$. The state (\ref{trsep}) can be considered
as the time-reversal-symmetric counterpart of the N\'eel state:
both of them have the same energy, the same correlation 
functions, etc., and they can not be distinguished by 
conventional experimental means (such as neutron scattering or 
NMR). Moreover, the energy of the TRS-state is very 
close to the energy of the exact AFM ground state: they differ 
by terms of order of $1/(zS)$ where $z$ is the coordination 
number \cite{neel} which is often large (e.g., $z=8$ for 
{\it bcc\/} lattice). For simplicity, 
below we consider only the case $S=1/2$, so that our treatment 
can be considered as formally rigorous in the limit $z\to\infty$, 
i.e. for infinite-dimensional space.
Note that the difference in energy between the states 
$|\Psi_M\rangle$ with $M$ close to $NS$ is of order $1/N$,
i.e. the ground state is highly (quasi)degenerate.

Below, to characterize the states of the antiferromagnet, it is
convenient to use the correlation functions
$
K^{\alpha\beta}({\bf q})=\mathop{\rm Tr} [\rho S^{\alpha}_{\bf q}
  S^{\beta}_{-\bf q}],
$
where $\rho$ is the density matrix. For the state
(\ref{trsep}) we have
\begin{eqnarray}
K^{+-}_i({\bf q}) &=& \langle S^+_{\bf q} S^-_{-\bf q} \rangle 
  = NS^2 \delta_{\bf q Q} + S/2, \\
 \nonumber
K^{zz}_i({\bf q}) &=& \langle S^z_{\bf q} S^z_{-\bf q} \rangle 
  = S/2,
\end{eqnarray}
where the subscript $i$ stands for the initial state,
and this state can be imagined as a state with all
the spins aligned antiferromagnetically along some AFM axis
(note the delta-functional singularity in $K^{+-}$),
but the axis itself is uniformly smeared in the $x$-$y$ plane,
see Fig.\ \ref{figep}a. Obviously, in this state the sublattices
are absent.

Now, suppose that we perform a local measurement of the spin 
$x$-projection at the site $0$. In reality, no conscious effort
is needed for that, since the electronic spin at every site
interacts with its environment (e.g., a nuclear spin at this
site), so the initial state of the electronic spin decoheres
\cite{zurek}, i.e., the environment acts as a measuring
apparatus. Decoherence of the single spin by its environment can
be described following the Ref.\ \cite{stamp}; in many realistic
cases this process is very fast, so we describe it
using von Neumann's theory, as a projection of the initial
state onto the eigenstates of the operator $s^x_0$. Therefore,
the system occurs in a mixed quantum state which is
described by the density matrix
\begin{eqnarray}
\rho_f &=& \left(|\Phi^+\rangle\langle\Phi^+|
  +|\Phi^-\rangle\langle\Phi^-|\right), \\
 \nonumber
|\Phi^{\pm}\rangle &=& W^{\pm} |\Psi_M\rangle = 
  \left(\frac 12 \pm s^x_0\right) |\Psi_M\rangle,
\end{eqnarray}
where, for $S=1/2$, $W^+$ and $W^-$ are the projectors onto the
eigenstates of $s^x_0$ with the eigenvalues $+1/2$ and $-1/2$ 
respectively. The function $|\Phi^+\rangle$
($|\Phi^-\rangle$) describes the state of the sample after
the measurement, provided that the measurement gave the 
result $+1/2$ ($-1/2$). Note that the norm 
of both $|\Phi^+\rangle$ and $|\Phi^-\rangle$ is $1/2$, i.e.
each state appears with the probability 1/2.
The calculations give the explicit form: 
\begin{eqnarray}
\label{wffin}
|\Phi^{\pm}\rangle &=& |\Phi^{\pm}_{\text{coh}}\rangle +
  |\Phi^{\pm}_{\text{inc}}\rangle, \\
\nonumber
|\Phi^{\pm}_{\text{coh}}\rangle &=& \frac 12 |\Psi_M\rangle 
  \pm \frac 12 \left( \alpha_M |\Psi_{M-1}\rangle + 
  \alpha_{M+1} |\Psi_{M+1}\rangle \right),\\
\nonumber
|\Phi^{\pm}_{\text{inc}}\rangle &=& \pm \frac 1{2\sqrt{N}}
  \sum_{{\bf q}\neq{\bf Q}} S^-_{\bf q} \left(|\Psi_M\rangle 
  - \gamma_M |\Psi_{M-2}\rangle\right),
\end{eqnarray}
where $|\Phi^{\pm}_{\text{coh}}\rangle$ represents
contributions from the states with ${\bf q}={\bf Q}$, while
$|\Phi^{\pm}_{\text{inc}}\rangle$ corresponds to 
the excitations with ${\bf q}\neq{\bf Q}$, and
$\alpha_M = \sqrt{M(N-M+1)/N^2}$, 
$\gamma_M = \sqrt{M(M-1)/[(N-M+1)(N-M+2)]}$.

The initial AFM ground state is invariant 
with respect to rotations in the $x$-$y$ plane, and so is the 
final state produced by the measurement:
\begin{eqnarray}
K^{+-}_f({\bf Q}) &=& K^{xx}_f({\bf Q}) + K^{yy}_f({\bf Q})
  = N/4,\\
 \nonumber
K^{--}_f({\bf Q}) &=& K^{xx}_f({\bf Q}) - K^{yy}_f({\bf Q})
  = 1/16,
\end{eqnarray}
for $N\gg 1$, where we took into account that $M=NS$ for the 
AFM ground state. Only the direction of the measured spin at
the site $j=0$ is fixed, while for the rest of the sample
the AFM axis in the $x$-$y$ plane is not fixed.
This is quite different from what is
predicted by the collective coordinate approach \cite{stamp}
or other models such as the classical $\sigma$-model.

However, in spite of the indeterminate AFM axis, sublattices
{\it are created\/} in the sample. Indeed, for every site
$j$ ($j\neq 0$):
\begin{equation}
\langle s^x_j\rangle_{\pm} = \frac{\langle\Phi^{\pm}| s^x_j 
  |\Phi^{\pm}\rangle}{\langle\Phi^{\pm}|\Phi^{\pm}\rangle} 
  = \pm \frac 14 \exp{(-i{\bf Qr}_j})
\end{equation}
i.e., for each result of the measurement, the $x$-component of
the spin at every site is nonzero, although twice smaller
than for the N\'eel state, and the spins are ordered
antiferromagnetically.
This state can be interpreted in quasiclassical terms 
as an AFM ``fan'' structure,
where due to the measurement the spins at the sites $j\neq 0$ are
localized essentially within the semicircles, but their direction
inside the semicircles is not fixed (see Fig.\ \ref{figep}b).
This picture can be confirmed by calculation of other quantities
(the amplitude $\langle S^x_{\bf Q}\rangle$ of the mode 
${\bf q}={\bf Q}$, single-spin density matrix, etc.).

Now, let us consider the case of the easy-axis anisotropy,
assuming that the easy axis is directed along the $x$-axis.
The TRS ansatz for the corresponding AFM ground state is
\cite{irkat}
\begin{equation}
|\Psi\rangle = \sum_{L=0}^{2NS} u_L |\Psi_{2L}\rangle,\quad
  \sum_{L=0}^{2NS} u^2_L = 1,
\end{equation}
where $u_L = \exp{[\phi(L)/2]}$, and $\phi(L)$ is some real
function, whose explicit form is immaterial in the limit
$N\to\infty$ provided that it has a well-pronounced maximum
in the neighborhood of $L_0=NS$. It has been shown \cite{irkat}
that this state corresponds to the AFM state without sublattices,
but with the AFM ordering axis directed along the $x$-axis.
As above, we consider the state of the system after the 
measurement made on the $x$-component of the spin at the site
$j=0$. The resulting density matrix is 
$\rho_f = |\Phi^+\rangle\langle\Phi^+| + 
|\Phi^-\rangle\langle\Phi^-|$, and in the thermodynamic limit
\begin{equation}
|\Phi^{\pm}\rangle = \frac 12 \sum_L
  u_L\left[ |\Psi_{2L}\rangle \pm \alpha_{2L+1}
  |\Psi_{2L+1}\rangle \pm \alpha_{2L}|\Psi_{2L-1}\rangle\right]
\end{equation}
while the excitations with ${\bf q}\neq {\bf Q}$ are negligible.
It can be checked that
the resulting state is, in fact, the N\'eel state: e.g.,
the $x$-component of the spin at the site $j$ is
$\langle s^x_j\rangle_{\pm} = \pm (1/2)\exp{(-i{\bf Qr}_j)}$,
exactly as in the N\'eel state (see Fig.\ \ref{figep}d). 

Therefore, for the antiferromagnet which is initially in the 
TRS-state, even the single site measurement can create 
sublattices over the whole sample. However, these sublattices have
the usual N\'eel form only for the easy-axis antiferromagnet. 
For the easy-plane antiferromagnet, the ``fan'' sublattices 
appear, where the AFM axis is indeterminate in the $x$-$y$ 
plane, and each spin is directed up or down only in average, 
being localized essentially in the semicircle. Our conclusions 
do not depend on the specific form of
the anisotropic term in the Hamiltonian, and are based only on
the use of the TRS-states for description of the AFM ground
state. This description is expected to be valid
for small anisotropy with accuracy of order $1/(zS)$. As a 
rule, this approximation works well \cite{mattis}.

To produce the N\'eel state in the easy-plane case, a 
macroscopically large number of
further measurements are necessary. Indeed, every measurement
admixes to the initial state $|\Psi_M\rangle$ the states
$|\Psi_{M\pm 1}\rangle$, so that to encompass all  
values of $M$ and fix the direction of the AFM axis, we need
of order of $N/2$ measurements. Informally, it can be understood
by taking into account the uncertainty relation for 
the $z$-projection of the total spin ${\cal S}_z$
and the angle $\phi$ of AFM axis in the $x$-$y$
plane: $\Delta{\cal S}_z \Delta\phi\sim 1$. The value of
$\phi$ becomes certain when $\Delta{\cal S}_z$ is
maximal. This situation is analogous to the case of 
Bose-Einstein condensate, where the well-defined phase is 
built up by increasing uncertainty in the number of 
particles \cite{bec}.

Nevertheless, appearance of the sublattices is not the whole
story. It has been shown \cite{decwav} that the local 
measurement in a distributed quantum system creates a
decoherence wave propagating in the system. In the 
antiferromagnet, the decoherence wave can be created only by 
the contributions
with ${\bf q}\neq {\bf Q}$: the excitations with 
${\bf q}={\bf Q}$ are static (have zero frequency). 
For example, in the easy-plane case 
$\langle s^x_j\rangle (t) = \pm \mathop{\rm Re} \langle
  s^+_j(t) s^-_0 \rangle$,
i.e.\ the local spin measurement induces the 
decoherence wave, and its dynamics 
is governed by the Green's function 
$G({\bf r}_{jk}, t) = \langle s^+_j(t) s^-_k \rangle 
  - \langle s^+_j s^-_k \rangle$.
For the easy-plane TRS-state $G({\bf r},t)$
can not be calculated exactly, but
estimates based on the magnon picture give 
$G({\bf r}, t) = (4N)^{-1} \sum_{\bf q} \exp{[i{\bf qr}
  - i\omega_{\bf q}t]}$, 
where $\omega_{\bf q}=\sqrt{(J_{\bf Q}-J_{\bf q})
(J_{\bf Q}-J_{{\bf q}-{\bf Q}})}$ is the frequency of
antiferromagnons. 
It is worth noting that the decoherence
wave appears only in the easy-plane case, where the localization
of the AFM axis is incomplete. In the easy-axis case the
AFM axis is localized completely but the incoherent contribution
in the wave function is negligible in the limit $N\to\infty$,
so the decoherence wave is absent.
This corroborates the conjecture \cite{decwav} that the 
decoherence wave is the result of 
incomplete entanglement between the entities comprising the
system. Indeed, in the easy-axis case the entanglement is 
complete, and measurement of the direction of one spin 
determines the
direction of every other spin in the sample. In the easy-plane
case, the entanglement is not so constraining, so that the
excitations with ${\bf q}\neq {\bf Q}$ emerge, making the
AFM axis indeterminate and giving rise to the decoherence wave.

Results of the present paper can have also experimental
importance. At present, there is a growing interest in 
antiferromagnetic systems of significant 
size which, nonetheless, demonstrate quantum coherence. 
Mention can be made of the single-molecule
antiferromagnets \cite{molafm} Fe$_{10}$, Fe$_6$, etc. Another
relevant class of experimentally interesting systems is the 
artificial spin rings made of
paramagnetic atoms to form a ``quantum corral'' \cite{corral}.
Noticeable AFM interactions can exist between the spins in this
system, while major sources of decoherence, nuclear spins and
conduction electrons \cite{stamp}, can be avoided, e.g., by
using iron atoms and silicon substrate at low temperatures.
The spin state of the atoms in the ring can be probed by 
scanning tunneling microscopy which can provide \cite{stm}
the spatial resolution of ca.\ 10 \AA\ and sensitivity 
of order of few Bohr's magnetons, or by magnetic resonance force 
microscopy \cite{mrfm}.
But development and analysis
of realistic experimental schemes constitutes a distinct
problem, requiring separate detailed study, and we do not 
discuss it here.

Summarizing, we have shown that the {\it local\/} measurement
in a distributed macroscopic antiferromagnet can create 
sublattices in the whole sample.
Therefore, considering the sublattices as a
result of decoherence, it is not necessary to presume a 
measurement embracing the whole 
macroscopic sample at once. We show that the result of 
measurement depends on the symmetry of the ground state.
For the easy-axis case the collective coordinate approach gives
essentially the complete picture, with both time-reversal and
rotational symmetries broken. For the easy-plane case,
unusual ``fan'' sublattices appear, i.e.\ only time-reversal
symmetry breaking takes place while rotational symmetry is 
preserved (in spite of high degeneracy of the ground state); 
moreover, in this case a decoherence wave
is generated by measurement. We conclude that both the certainty 
of the final state
and the appearance of the decoherence wave are governed by the
degree of entanglement of spins in the system. 

This work was partially carried out at the Ames Laboratory, which 
is operated for the U.\ S.\ Department of Energy by Iowa State 
University under Contract No.\ W-7405-82 and was supported by 
the Director for Energy Research, Office of Basic Energy Sciences 
of the U.\ S.\ Department of Energy.

\begin{figure}
\caption{Sketch of the spin arrangement. Easy plane case:
(a) before measurement spins are antiferromagnetically ordered,
but sublattices are absent and the total AFM axis is not fixed;
(b) after measurement the ``fan'' sublattices emerge but AFM axis
is not fixed. Easy axis case: (c) before measurement spins are
directed along $x$-axis but sublattices are absent; (d) after
measurement the N\'eel state appears.}
\label{figep}
\end{figure}


\begin{references}
\bibitem[*]{mik} Permanent address: Institute of Metal Physics, 
  Ekaterinburg 620219\, Russia.
\bibitem{neel} S. V. Vonsovsky, {\it Magnetism\/}, vol. 2 (Johh 
  Wiley \& Sons, New York, 1974) and references therein.
\bibitem{astrov} D. N. Astrov, Sov. Phys. JETP {\bf 13},
  729 (1961); 
  D. N. Astrov, N. B. Ermakov, 
  A. S. Borovik-Romanov, E. G. Kolevatov, and 
  V. I. Nizhankovskii, JETP Lett. {\bf 63}, 745 (1996).
\bibitem{pwa} P. W. Anderson, {\it Basic Notions of Condensed 
  Matter Physics\/} (Benjamin/Cummings, Menlo Park, 1984).
\bibitem{zurek} W. H. Zurek, Phys. Rev. D {\bf 24}, 1516 (1981);
  {\bf 26}, 1862 (1982); E. Joos and H. D. Zeh, Z. Phys. B 
  {\bf 59}, 223 (1985); A. Leggett {\it et al\/}., Rev. Mod. 
  Phys. {\bf 59}, 1 (1987).
\bibitem{vonneu} J. von Neumann, {\it Mathematical
  Foundations of Quantum Mechanics\/} (Princeton, Princeton
  University Press, 1955).
\bibitem{stamp} A. Garg, in {\it Quantum Tunneling of 
  Magnetization --- QTM'94\/}, edited by L. Gunther and 
  B. Barbara (Kluwer Academic Publishers, Dordrecht, 1995);
  N. V. Prokof'ev and P. C. E. Stamp, {\it ibid.\/};
  N. V. Prokof'ev and P. C. E. Stamp, Rep. Prog. Phys.
  {\bf 63}, 669 (2000).
\bibitem{twostate} The state $|1\rangle$ can be the
  state where all the spins from the sublattice $A$ are directed
  up and all the spins from the sublattice $B$ are directed down,
  while in the state $|2\rangle$ the spins have opposite
  directions (down for the sublattice $A$ and up for the
  sublattice $B$).
\bibitem{irkat} V. Yu. Irkhin and M. I. Katsnelson, Z. Phys. B 
  {\bf 62}, 201 (1986).
\bibitem{decwav} M. I. Katsnelson, V. V. Dobrovitski, and 
  B. N. Harmon, Phys. Rev. A {\bf 62}, 022118 (2000).
\bibitem{mattis} D. Mattis, {\it The Theory of Magnetism\/}
  (Harper \& Row, New York, 1965).
\bibitem{bec} J. I. Cirac, C. W. Gardiner, M. Naraschewski, and
  P. Zoller, Phys. Rev. A {\bf 54}, R 3714 (1997).
\bibitem{molafm} D. Gatteschi, A. Caneschi, L. Pardi and 
  R. Sessoli, Science {\bf 265}, 1054 (1994).
\bibitem{corral} H. C. Manoharan, C. P. Lutz, and D. M. Eigler, 
  Nature {\bf 403}, 512 (2000).
\bibitem{stm} Y. Manassen, R. J. Hamers, J. E. Demuth, and A. J.
  Castellano, Jr., Phys. Rev. Lett. {\bf 62}, 2531 (1989).
\bibitem{mrfm} O. Rugar, C. S. Yannoni, and J. A. Sidles, Nature
  {\bf 360}, 563 (1992); G. P. Berman and V. I. Tsifrinovich,
  Phys. Rev. B {\bf 61}, 3524 (2000).
\end{references}
\end{document}